\documentclass[12pt,preprint]{aastex}

\shorttitle{Short Title} \shortauthors{Tang \& Fan}

\begin{document}


\title{Effects of the complex mass distribution of dark matter halos 
on weak lensing cluster surveys}


\author{J. Y. Tang and Z. H. Fan}
\affil{Department of Astronomy, Peking University,
    Beijing 100871, China}
\email{tangjy@bac.pku.edu.cn,fan@bac.pku.edu.cn}


%\altaffiltext{1}{Visiting Astronomer, Cerro Tololo Inter-American Observatory.
%CTIO is operated by AURA, Inc.\ under contract to the National Science
%\altaffiltext{2}{Society of Fellows, Harvard University.}
%\altaffiltext{3}{present address: Center for Astrophysics,
%    60 Garden Street, Cambridge, MA 02138}
%\altaffiltext{4}{Visiting Programmer, Space Telescope Science Institute}
%\altaffiltext{5}{Patron, Alonso's Bar and Grill}


%\begin{abstract}
%This is a preliminary report on surface photometry of the major
%fraction of known globular clusters, to see which of them show the signs
%of a collapsed core.
%We also explore some diversionary mathematics and recreational tables.
%\end{abstract}
\begin{abstract}
Gravitational lensing effects arise from the light ray 
deflection by all of the mass distribution along the line of sight.
It is then expected that weak lensing cluster
surveys can provide us true mass-selected cluster samples. With
numerical simulations, we analyze the correspondence
between peaks in the lensing convergence $\kappa$-map and dark matter halos.
Particularly we emphasize the difference between the peak $\kappa$ value
expected from a dark matter halo modeled as an isolated and spherical one,
which exhibits a one-to-one correspondence with the halo mass 
at a given redshift, and that of the associated $\kappa$-peak from 
simulations. For halos with the same expected $\kappa$, 
their corresponding peak signals in the $\kappa$-map present a 
wide dispersion. At an angular smoothing scale of $\theta_G=1\hbox{ arcmin}$,
our study shows that for relatively
large clusters, the complex mass distribution of individual clusters 
is the main reason for the dispersion. The projection effect
of uncorrelated structures does not play significant roles. 
The triaxiality of dark matter halos accounts for a large
part of the dispersion, especially for the tail at high $\kappa$ side. 
Thus lensing-selected clusters are not
really mass-selected. To better predict $\kappa$-selected cluster 
abundance for a cosmological model, one has to take into account the triaxial 
mass distribution of dark matter halos. On the other hand, for a significant number
of clusters, their mass distribution is even more complex than that described by the
triaxial model. Our analyses find that large substructures
affect the identification of lensing clusters considerably. They 
could show up as separate peaks in the $\kappa$-map, and 
cause a mis-association of the whole cluster with a peak resulted 
only from a large substructure. The lower-end dispersion of 
$\kappa$ is attributed mostly to this substructure effect.
For $\theta_G=2\hbox{ arcmin}$, the projection effect can be significant
and contributes to the dispersion at both high and low $\kappa$ ends.
\end{abstract}




%\section{Introduction}
%
%A focal problem today in the dynamics of globular clusters is
%core collapse.  It has been predicted by theory
%for decades \citep{hen61,lyn68,spi85}, but
%observation has been less alert to the phenomenon. For many years the
%central brightness peak in M15 \citep{kin75,new78}
%seemed a unique anomaly.  Then \citet{aur82} suggested a central peak
%in NGC 6397, and a limited photographic survey of ours \citep[Paper I]{djo84}
%found three more cases, including NGC 6624, whose
%sharp center had often been remarked on \citep{can78}.
%ii
%i$
%i$
\section{Introduction}
Because they are directly associated with the mass distribution of 
the universe, gravitational lensing effects are powerful probes of spatial
structures of the dark matter. Strong lensing phenomena, such as
multiple images of background quasars and giant arcs, have been used to 
constrain inner mass profiles of lensing galaxies and clusters
of galaxies (e.g., Gavazzi et al. 2003; Bartelmann \& Meneghetti 2004;
Ma 2003; Zhang 2004). Weak lensing
effects, on the other hand, enable us to study 
mass distributions of clusters of galaxies out to large radii
(e.g., Bartelmann \& Schneider 2001). 
Cosmic shears, coherent shape distortions of background galaxies 
induced by large-scale structures in the universe, provide us
a promising means to map out the dark matter distribution of the
universe (e.g., Tereno et al. 2005; Van Waerbeke 2005).

Of many important studies on lensing effects, the aspects of
weak lensing cluster surveys attract more and more attention
(e.g., Reblinsky \& Bartelmann 1999; White et al. 2002; 
Padmanabhan et al. 2003; Hamana et al. 2004; Haiman et al. 2004). 
Clusters of galaxies are the largest virialized structures 
in the present universe. Their formation and evolution
are sensitive to cosmologies, and therefore can be used
to constrain different cosmological parameters, such as 
$\sigma_8$, $\Omega_m$ and the equation of state of dark energy,
where $\sigma_8$ is the rms of the extrapolated linear density fluctuation
smoothed over $8h^{-1}\hbox{Mpc}$, and $\Omega_m$ is the present
matter density in units of the critical density of the universe
(e.g., Bahcall \& Bode 2003; Fan \& Chiueh 2001; Fan \& Wu 2003; 
Haiman et al. 2001) .
There are different ways finding clusters. The optical
identification based on the concentration of galaxies
suffers severe projection effects. X-ray and
Sunyaev-Zel'dovich (SZ) effect are associated with the intracluster gas,
and have been used extensively in cluster studies (e.g., Rosati et al. 2002;
Carlstrom et al. 2002). However, most of the theoretical studies 
concern the abundance of clusters in terms of their masses 
(e.g., Press \& Schechter 1974; Sheth \& Tormen 1999; Jenkins et al. 2001), 
therefore it is crucial to get
reliable relations between different survey observables and clusters' mass.
The properties of intracluster gas are affected significantly by gas physics,
which we have not fully understood yet.
Thus there are large uncertainties in relating
X-ray and SZ effect with the total mass of a cluster. On the other hand,
lensing effects of a cluster are determined fully by its mass distribution,
and therefore clean mass-selected cluster samples are
expected from weak lensing cluster surveys.

However, weak lensing surveys have their own complications. Lensing
effects are associated with the mass distribution between sources
and observers, and thus the lensing signal of a cluster can be
contaminated by other structures along the line of sight.
The intrinsic ellipticities of source galaxies can pollute the 
lensing map and lower the efficiency of cluster detections considerably.
Besides and more intrinsically, clusters themselves generally have complex mass 
distributions, and their lensing effects can be affected by 
different factors in addition to the total mass. 

Therefore for extracting cosmological information from a sample 
of lensing-selected clusters, three main theoretical issues need to 
be carefully studied. Firstly the lensing effects from clusters
must be understood thoroughly. Secondly the significance of projection effects 
along the line of sights should be estimated.  
Thirdly the noise due to the intrinsic asphericity of source galaxies
should be treated properly.  
It is important to realize that the existence of noise can affect the 
detection of clusters considerably. Numerical studies (Hamana et al. 2004;
White et al. 2002) showed that the presence of noise reduces the
efficiency of cluster detection significantly. Van Waerbeke (2000) investigated
the properties of noise induced by the intrinsic ellipticities of 
source galaxies. He found that in the weak lensing regime,
the lensing signal and the noise are largely uncorrelated if the 
smoothed convergence $\kappa$ is considered. 
Furthermore, to a very good approximation,
the noise can be described as a two-dimensional Gaussian random field 
with the noise correlation introduced only through smoothing procedures. 
Then the technique of Bardeen et al. (Bardeen et al. 1986) can be used 
to calculate the number of peaks resulted purely from the noise. This provides
us a possible way to estimate the contamination of noise on the 
abundance of lensing-detected clusters. The presence
of noise also affects the height of peaks from real clusters.
With numerical simulations, Hamana et al. (2004) tried to correct
this effect empirically. In our future work, we will study the noise
in great detail with the effort to establish a model to describe
its effects on weak lensing cluster surveys. 
With the hope that this is achievable, we address in this paper the
first two issues with the emphasis on the effects of the complex
mass distribution of clusters themselves.

Even for isolated clusters without any projection effect and
without any noise, their lensing effects 
cannot be fully determined by their mass. Thus lensing-selected clusters
cannot be truly mass-selected. 
%Notice that the efficiency and completeness depend on the 
%base cluster sample we choose to compare with. The base sample
%should consist of clusters detectable by lensing effects under
%ideal conditions that only clusters contribute
%to lensing signals without any projection effects and without
%any noises. Thus in order to theoretically predict the yield
%of a lensing survey for a specific cosmological
%model, it is important to first understand the ideal selection 
%function in detail. 
%White et al. (2002) and Padmanabhan et al. (2003) 
%studied the efficiency and completeness of weak lensing cluster detection
%considering clusters above a fixed mass limit. 
Hamana et al. (2004) adopted the spherical Navarro-Frenk-White (NFW)
(Navarro et al. 1996, 1997) density profile for a cluster to relate
its smoothed peak $\kappa$ value with its total mass. Given a
detection limit on $\kappa$, they then obtained an ideal mass selection 
function with the redshift-dependent lower limit derived from the 
limit of $\kappa$.
The essence of their model is still that there is a one-to-one correspondence
between the peak $\kappa$ of a cluster (at a given redshift)
and its total mass. However, the complicated mass distribution of a cluster
can make its lensing effect much more complex than that predicted solely
by its total mass. In this paper, we compare the peak $\kappa$ values of 
clusters from numerical simulations with the results by 
modeling the dark matter halos of clusters as isolated and
spherical ones. Large differences in $\kappa$ exist for those clusters
with the same expected $\kappa$ from the NFW-profile analysis.
Our investigation further finds
that the triaxiality of the mass distribution of clusters and large 
substructures in them contribute to this dispersion significantly.

The paper is organized as follows. In \S2, we describe the lensing
simulations and the cluster identification scheme from lensing
convergence $\kappa$-map. In \S3, we present the results of
our analyses. Summary and discussion are included in \S4.

%In the Appendix, we introduce the calculation of the eccentricity for the other
%observations and the calculation of the parameter $C/K$.

\section{Weak lensing simulations}
%\subsection{Gas density distribution}

In this paper, we consider the concordance cosmological model
with $\Omega_m=0.3$, $\Omega_{\Lambda}=0.7$, $\Gamma=0.2$
and $\sigma_8=0.9$, where $\Omega_m$ and $\Omega_{\Lambda}$
are the present matter density and the energy density of 
the cosmological constant
in units of the critical density of the universe, $\Gamma$ is the
shape parameter of the spectrum of the linear density fluctuations,
and $\sigma_8$ is the rms of the extrapolated linear density 
perturbation smoothed over $8h^{-1}\hbox{Mpc}$ with $h$ being the
present Hubble constant in units of $100\hbox{ kms}^{-1}\hbox{ Mpc}^{-1}$. 

The simulations we use (Jing \& Suto 1998) have the box size of 
$100h^{-1}\hbox{Mpc}$, and $256^3$ particles. The
mass of each particle is $5\times 10^{9}h^{-1}\hbox{M}_\odot$.
The softening length is $39h^{-1}\hbox{kpc}$. A typical cluster
has a mass of $\sim 10^{14}\hbox{M}_\odot$ and a size of $\sim \hbox{Mpc}$,
thus both the mass resolution and the force resolution are
good enough for our studies. There are three simulation runs
with different realizations of the initial conditions. 
For each simulation, it evolves from redshift $z=72$ to $z=0$
with $1200$ time steps equally spaced in terms of
the cosmological scale factor. There are total $60$ outputs 
with $14$ of them between $z=1$ and $z=0$.
Dark matter halos are identified with the FOF algorithm with the linking length 
$b=0.2$ in units of the average separation of particles. 

In this paper, we focus on the weak lensing regime. In this limit, the shear
$\gamma$ can be directly obtained from image distortions through
$\epsilon^{(I)}\approx \gamma + \epsilon^{(S)}$ where $\epsilon$ is the
ellipticity defined in the complex plane and $I$ and $S$ stand for image and
source, respectively. Because both the convergence $\kappa$ 
and the shear $\gamma$
are determined by the lensing potential, $\kappa$ can be estimated 
from $\gamma$ (e.g., Kaiser 1998). Then we have 
$\kappa_n(\theta)=\kappa(\theta)+n(\theta)$
where $\kappa$ is the true lensing convergence, and $n$ is the noise part 
associated with the intrinsic ellipticity of source galaxies $\epsilon^{(S)}$
(e.g., Van Waerbeke 2000). Detailed analysis by Van Waerbeke (2000)
shows that for the suitably smoothed $\kappa_n$,
the lensing signal and the noise are uncorrelated in the weak lensing limit,
and the noise can be well approximated by a two-dimensional Gaussian 
field. In his analysis, 
Van Waerbeke ignored the spatial clustering and the correlation
of intrinsic ellipticities of source galaxies on $arcmin$ scales.
Therefore the smoothed noise is equivalent to the convolution
of a point process by a smoothing window. According to the central limit
theorem, the statistics of the smoothed noise is approaching 
Gaussian when the number of source galaxies contained in the smoothing 
window is large enough. Although observations and numerical simulations 
indicate that the simplification mentioned above may be indeed valid
on $arcmin$ scales for source galaxies at $z\sim 1$ 
(e.g., Van Waerbeke 2000), the statistics of the noise
can deviate from Gaussian if there are significant correlations 
in the spatial distribution or in the intrinsic ellipticity of source galaxies. 
In our future studies, we will investigate in detail 
the noise properties with these correlations included. 

It is very true that the existence of
noise affects the identification of clusters significantly.
On the other hand, for theoretical studies, it is fundamentally important to 
first understand the complexity of true lensing signals.
The knowledge of this part forms the base for further investigations. 
Thus in this paper, we focus on true lensing signals 
with the emphasis on the influence of the complicated mass distribution
of clusters and the projection effect. Therefore we
mainly study noise-free $\kappa$ maps. The effects of 
noise on lensing cluster surveys will be carefully analyzed
in our follow-up studies.

In theoretical studies of lensing effects, generally one obtains the deflection 
angle $\vec \alpha$ through ray-tracing simulations (e.g., Jain et al. 2000). 
Then the shear matrix is calculated by 
$\Phi_{ij}=\partial \alpha_{i}/\partial \beta_{j}$
with $\vec \beta$ the source angular position. The convergence $\kappa$ and
the shear $\gamma$ can be derived from $\Phi_{ij}$. 
%Numerical simulations have shown that in the weak lensing limit,
%$\gamma$ and $\kappa$ carry the same information of
%the mass distribution neglecting intrinsic ellipticities of source galaxies
%(e.g., Hamana et al. 2001). Because $\kappa$ is a weighted integral of 
%the density fluctuations along the path of photons, it is often the quantity
%studied in theoretical analyses. 
In terms of denisty perturbations, to a very good approximation, 
we have (Jain et al. 2000) 
\begin{equation}
\kappa=\frac{3H_0^2}{2c^2}\Omega_m\int_0^{\chi}g 
{\delta \over a} d\chi^{\prime},
\end{equation}
where $g(\chi^{\prime},\chi)=r(\chi^{\prime})r(\chi-\chi^{\prime})/r(\chi)$
with $r(\chi)$ being the comoving angular diameter distance,
$\delta$ is the density fluctuation at the actual position of photons, 
$H_0$ is the Hubble constant and $c$ is the speed of light.
%Numerically, multiple-lens-plane technique is often used in 
%generating $\kappa$ maps. To get the actual position of photons at 
%each lens plane, ray-tracing simulations are generally needed (e.g., Jain
%et al. 2000). 
In the weak lensing limit,
the Born approximation is valid and the integration in eq. (1)
can be approximately performed along the unperturbed light path.
We will see in \S3.1 that our study 
presented in this paper concerns $\kappa$ values of order $\sim 0.1$ 
for the Gaussian smoothing radius of $1\hbox { arcmin}$. Therefore the weak 
lensing approximation is well applicable. 
In this limit, $\kappa$ maps can be constructed from stacking different 
slices of mass distribution between sources and the observer together 
without intensive ray-tracing simulations. But we should keep in mind that
at sub-smoothing scales, the $\kappa$ value can be much higher.
The full lensing treatment should be carried out when studying
lensing effects at these smaller scales. 
Our method of stacking in the weak lensing limit is detailed below.

%For lensing effects, as many others, we consider the quantity 
%of convergence $\kappa$, which is the dimensionless surface mass
%density. On the other hand, it is the shear $\gamma$
%that is directly linked to observables. Thus shear maps
%resemble more real observations than $\kappa$-maps. In the weak lensing regime,
%however, $\gamma$ and $\kappa$ carry the same information of 
%the mass distribution neglecting intrinsic ellipticities of source galaxies
%(e.g., Hamana et al. 2001).
%Non-correlated intrinsic ellipticities of galaxies can be taken into
%account in the convergence map by modeling them as random Gaussian noise
%(van Waerbeke 2000; White et al. 2002; Hamana et al. 2004).
%The intrinsic alignments of source galaxies are difficult to model
%in $\kappa$-maps without ray-tracing simulations. 
%But for high redshift source galaxies (e.g., 
%$z_s\sim 1$), we expect that this contamination is negligible. 

To generate a $\kappa$ map, we use a multiple-lens-plane algorithm. 
Many other similar algorithms choose line-of-sight
directions to be parallel to the box edges ($z$-direction, for example)
(e.g., White et al. 2002; Hamana et al. 2001). To diminish the
effects of periodicity, the simulation boxes are randomly shifted
when they are stacked together. In the algorithm we apply, the stacking is
in a regular way without any shifts, but the line-of-sight
directions are chosen randomly. 
The effects of periodicity are minimal if we avoid
special directions, such as $30^{o}$, $45^{o}$ and
$60^{o}$ (Jing 2004, private communications). Specifically, 
we fix the source redshift at $z_s=1$. The observer is
at $z=0$ and locates at the position $(x,y,z)=(0.5,0.5,0.5)$.
The comoving radial distance between $z_s=1$ and $z=0$ is
divided into $40$ equal sections, whose corresponding redshifts
$z$ or the scale factors $a$ can be obtained. Then each slice
is filled by the outputting simulation boxes with their $a$ value 
closest to the $a$ value of the slice. 
The projected mass distribution in that slice is calculated
on $1024\times 1024$ grids for a $2\times 2\hbox{ deg}^2$ map. 
Thus along a line-of-sight direction, the convergence $\kappa$
is calculated through
\begin{equation}
\kappa=\frac{3H_0^2}{2c^2}\Omega_m\sum_{i=1}^{N}{\delta_{i}\over a_i}\frac{r_{\rm ds}r_{\rm d}}{r_{\rm s}}\Delta_i,
\end{equation}
where $i$ denotes different slices, $\delta_i$ is the three-dimensional
density perturbation, $a_i$ is the scale factor, $\Delta_i$ is the 
comoving thickness of a slice, 
$r_s$ and $r_d$ are the comoving angular diameter distances to the source
and to the lens, respectively, $r_{ds}$ is the comoving angular diameter
distance between the lens and the source.

%For the noise from intrinsic ellipticities of source galaxies,
%it is usually modeled as a two-dimensional Gaussian field based on 
%the study of Van Waerbeke (2000) and is added into the $\kappa$ map
%(e.g., White et al. 2002; Hamana et al. 2004). In our future studies,
%we will investigate the noise properties In this paper, 
%we focus on the complexity of the real lensing effects due to the
%mass distribution of clusters of galaxies and the projection effects.

%Besides observational effects, the intrinsic ellipticities
%of source galaxies introduce noises to the lensing signals. 
%To lower the noise level, a smoothing procedure is usually
%applied to the lensing map. 
%Since our focus is to investigate the ideal cluster selection
%function but not the effects of noises, we do not add in noises
%explicitly in our $\kappa$ maps. However, we smooth our noise-free
%maps with a Gaussian window function of the form
We smooth $\kappa$ maps with a Gaussian window function of the form
\begin{equation}
    W_G(\theta)=\frac{1}{\pi \theta^2_{G}}\exp(-\frac{\theta^2}{\theta_{G}^2}),
    \end{equation}
where $\theta_G$ is the smoothing angular scale. 
%In most of our
%studies, we take $\theta_G=1\hbox{ arcmin}$ (Hamana et al. 2004).
%We also do analyses with $\theta_G=2\hbox{ arcmin}$. The effects of
%different smoothing scales will be discussed 
Figure 1 shows a smoothed $\kappa$-map of $2\times 2\hbox{ deg}^2$
with $\theta_G=1\hbox{ arcmin}$. 
The grey-scale indicator shown in the right is in terms of the 
signal-to-noise ratio explained in the next section.
The squares denote dark matter halos found in the corresponding directions
with mass $M\geq 10^{14}h^{-1}\hbox{ M}_{\odot}$. Qualitatively, we do see 
good associations between peaks in the $\kappa$-map and
massive dark matter halos. 

To identify a match between a peak and
a halo quantitatively, we adopt the algorithm of Hamana et al. (2004). 
Specifically, the matching is carried out in two directions: whether a
peak has a corresponding halo and whether a halo has an associated
peak in the $\kappa$-map. Because it is not expected that
the position of a peak coincides exactly with the central position of 
its corresponding halo, the matched pair candidate is searched
within a radius of $2.88\hbox{ arcmin}$ around a peak or 
the center of a halo. The specific number $2.88\hbox{ arcmin}$ is chosen 
to be the same as that of Hamana et al. (2004), and it corresponds 
to $24$ pixels in our $\kappa$ maps. The closest one is identified 
as the primary match. Then there are five different peak-halo matching classes:
double primary match, missing halo, false peak, secondary peak
and secondary halo. A peak-halo pair is classified as 
a double primary match if both
the peak and the halo are their corresponding primary matches.
A missing halo is that there is no matched peak in the searching area
around the halo. A false peak is that there is no associated halo
around the peak. If a halo has a matched peak that has 
its own primary matched halo, this halo is classified as a secondary 
halo. A secondary peak has an associated halo, but is not the 
primary peak of that halo (Hamana et al. 2004). 
%Our investigation
%focuses on the selection function but not the completeness and
%efficiency, therefore we only consider the class of double primary match.
Our investigation focuses on the class of double primary match.

We do $2\times 2\hbox{ deg}^2$ surveys toward $720$ directions for
each simulation run. Of them, we discard those that have nearby halos
($z<0.1$) more massive than $5\times 10^{13}h^{-1}\hbox{ M}_{\odot}$. 
The final number of surveys used in our analysis is $428$, $487$, 
and $401$ for the three simulations, respectively.

\section{Results}

\subsection{NFW density profiles}

The NFW density profile of a dark matter halo is described by 
\begin{equation} 
\rho(x)=\frac{\rho_s}{(r/r_s)(1+r/r_s)^2},
\end{equation}
where $r_s$ and $\rho_s$ are the characteristic scale and density
of the halo, respectively. Given the virial mass $M_{vir}$ of the halo, 
$r_s$ can be calculated through the fitting formula
\begin{equation}
c_n={c_{n*}\over 1+z}\bigg ({M\over 10^{14}h^{-1}M_{\odot}}\bigg )^{-0.13},
\end{equation}
where the concentration parameter $c_n=r_{vir}/r_s$ with $r_{vir}$ 
the virial radius of the halo, 
and $c_{n*}=8$ for the cosmological model considered here. Then $\rho_s$
can be obtained by 
\begin{equation}
M_{vir}={4\pi\rho_s r_{vir}^3\over c_n^3}\bigg [ \log(1+c_n)-{c_n\over 1+c_n}\bigg ].
\end{equation}
Thus the density profile of a halo is fully determined by its mass.
Cutting off the mass distribution of a halo at its virial radius, the surface
mass density can be written as (e.g., Hamana et al. 2004)
\begin{equation}
\Sigma(x)=2\rho_s r_s f(x),
\end{equation}
where $x=r/r_s$ and
\begin{equation}
f(x)=\left\{%
\begin{array}{ll}
    {\sqrt{c_n^2-x^2}\over (x^2-1)(1+c_n)}+{1\over (1-x^2)^{3/2}}\hbox{arccosh}{x^2+c_n\over x(1+c_n)}, & \hbox{(x$< $1),} \\
    {\sqrt{c_n^2-1}\over 3(1+c_n)}\bigg (1+{1\over 1+c_n}\bigg ), & \hbox{(x$=$1),} \\
    {\sqrt{c_n^2-x^2}\over (x^2-1)(1+c_n)}+{1\over (1-x^2)^{3/2}}\hbox{arccos}{x^2+c_n\over x(1+c_n)}, & \hbox{ (1$<$ x$\leq$ $c_n$),} \\
    0, & \hbox{(x$>$ $c_n$)}.
	    \end{array}%
	    \right.
	    \end{equation}
The convergence $\kappa$ is then
\begin{equation}
\kappa={\Sigma(x)\over \Sigma_{cr}}=\kappa_s f(x),
\end{equation}
where 
\begin{equation}
\Sigma_{cr}={c^2\over 4\pi G}{D_s\over D_d D_{ds}},
\end{equation}
with $D_s$, $D_d$ and $D_{ds}$ are the angular diameter distances
to the source, to the lens, and from the lens to the source, respectively,
and
\begin{equation}
\kappa_s=2\rho_s r_s \Sigma_{cr}^{-1}.
\end{equation}
Then one can obtain the smoothed central $\kappa$ from a dark matter halo
by
\begin{equation}
\kappa_0=\int 2\pi \kappa(\theta / \theta_s)W_G(\theta,\theta_G) 
\theta d\theta.
\end{equation}
Thus for a NFW spherical dark matter halo, there is
a one-to-one correspondence between its $\kappa_0$ and its mass $M_{vir}$.
%Here we investigate if the spherical NFW approximation
%can describe well the lensing signal from a halo in simulations. 

We consider double-primary-matched halo-peak pairs. For each halo, from its
mass and redshift, we calculate the expected $\kappa_{0NFW}$, and compare
it with the real $\kappa$ value of the corresponding peak. 
Similar to other studies (e.g., White et al. 2002; Hamana et al. 2004),
we use the signal-to-noise ratio $\nu=\kappa_0/\sigma_{noise}$ to represent
the height of a peak with $\sigma_{noise}$ given by 
\begin{equation}
\sigma_{noise}^2={\sigma_{\epsilon}^2\over 2}{1\over 2\pi\theta_G^2 n_g},
\end{equation}
where $\sigma_{\epsilon}$ is the rms of the intrinsic ellipticity of source
galaxies, $n_g$ is the surface number density of source galaxies. Following
Hamana et al. (2004), we take $\sigma_{\epsilon}=0.4$ and 
$n_g=30\hbox{ arcmin}^{-2}$. In most of our analyses, we use 
$\theta_G=1\hbox{ arcmin}$. The corresponding $\sigma_{noise}=0.02$.
For a typical cluster of $M\sim 10^{14}\hbox{M}_{\odot}$,
the peak $\kappa \sim 0.1$ and thus $\nu \sim 5$ for
$\theta_G=1\hbox{ arcmin}$. Therefore it is appropriate to choose $\theta_G$
in the $arcmin$ scale for clusters of galaxies (Hamana et al. 2004). 
To illustrate the effect of using a different $\theta_G$, 
we will also present results with $\theta_G=2\hbox{ arcmin}$, in which 
$\sigma_{noise}=0.005$.

In Figure 2, we show the scatter plots of $\nu_{NFW}$ and $\nu_{peak}$ 
for the three simulation runs,
where $\nu_{NFW}$ stands for the expected central signal-to-noise
value for a halo modeled as a spherical NFW one, and $\nu_{peak}$  
is the halo's matched peak value from $\kappa$-maps. We only
consider halos with $M\ge 5\times 10^{13}h^{-1}\hbox{ M}_{\odot}$ and
peaks with $\nu_{peak}\ge 3$. From the plots, we do see
a correlation between $\nu_{NFW}$ and $\nu_{peak}$, but the 
dispersion is rather large. Quantitatively, we find that the dispersion
$\sigma \sim 1.2, 1.3 $, and $1.4$ for $\nu_{NFW}=4.5, 5$ and
$6$, respectively. We also note that
the average $\nu_{NFW}-\nu_{peak}$ is slightly off zero, with
$\nu_{NFW}-\nu_{peak} \sim -0.3$. Hamana et al. (2004) found 
a positive bias with $\nu_{NFW}-\nu_{peak}\sim 0.24$. It is 
likely that the small bias arises from simulation to simulation variances.
On the other hand, the large dispersion indicates 
that for a cosmological model, the lensing-selected cluster abundance can 
deviate significantly from that predicted based on the spherical 
NFW description of cluster profiles. Therefore the NFW approximation
can introduce large systematic errors in extracting cosmological 
parameters from lensing cluster surveys. 

\subsection{Triaxial dark matter halos}

After a rough calculation, Hamana et al. (2004) 
attributed the dispersion to the statistical distribution of the
concentration parameter of the NFW density profile,
and the projection effects of uncorrelated structures along the line of sight. 
Here we will study in detail the reasons of the dispersion, with
particular attention paid to the triaxiality of dark matter halos.

With high resolution simulations, Jing and Suto (2002) present
a NFW-like triaxial density profile for dark matter halos, which is
given by
\begin{equation}
\rho(R)=\frac{\rho_{cr}(z)\delta_{ce}}{(R/R_0)^{\alpha}(1+R/R_0)^{3-\alpha}},
\end{equation}
where $\rho_{cr}(z)$ is the critical matter density of the universe
at redshift $z$ and
\begin{equation}
R^2=\frac{x^{\prime 2}}{c_x^2}+\frac{y^{\prime 2}}{c_y^2}+z^{{\prime} 2},
\end{equation}
with the axial ratios $c_x \leq c_y \leq 1$. The characteristic scale $R_0$ is
related to the triaxial concentration parameter $c_e=R_e/R_0$ with 
$R_e\approx 0.45 R_{vir}$ defined in Jing and Suto (2002).
Given the mass of a halo, the average $c_x$ and $c_y$, and $c_e$,
as well as their statistical distributions are derived explicitly from 
simulation results (Jing \& Suto 2002).

The effects of the triaxiality on strong lensing results, such as
arc statistics and the probability of large-separation multiple images,
have been studied in detail (e.g., Oguri et al. 2003; Oguri \& Keeton 2004).
The associated asphericity of the intracluster gas, revealed 
by X-ray and Sunyaev-Zeldovich effect observations, has also been
investigated (e.g., Lee \& Suto 2003, 2004; Wang \& Fan 2004).
Here we analyze the dispersion between $\nu_{peak}$ and $\nu_{NFW}$
caused by the aspherical mass distribution of dark matter halos. 

The $\kappa$ profile of a triaxial dark matter halo can be
analytically written as (Oguri et al. 2003)
\begin{equation}
\kappa(\zeta)=\frac{2\delta_{ce}\rho_{cr}R_0}{\sqrt{g}\Sigma_{cr}}f(\zeta),
\end{equation}
where the factor $g$ is defined as
\begin{equation}
g=\sin^2\theta(\frac{1}{c_x^2}\cos^2\phi+\frac{1}{c_y^2}\sin^2\phi)+
\cos^2\theta,
\end{equation} 
with $(\theta, \phi)$ the polar coordinates of the line-of-sight direction.
The factor $f$ is the same as that in equation (8) but with the 
variable $\zeta$ labeling the elliptical contours of the projected 
surface mass distribution. Specifically,
\begin{equation}
\zeta^2={1\over g}(Ax^2+Bxy+Cy^2),
\end{equation}
where we have assumed that the lensing plane is described by $(x,y)$
coordinates, and 
\begin{equation}
A=\cos^2\theta \bigg ({1\over c_x^2} \sin^2\phi +{1\over c_y^2} \cos^2\phi
\bigg )+{1\over c_x^2 c_y^2}\sin^2\theta,
\end{equation}

\begin{equation}
B=\cos \theta \sin 2\phi\bigg ({1\over c_x^2}-{1\over c_y^2}\bigg ),
\end{equation}
and
\begin{equation}
C={1\over c_y^2}\sin^2\phi+{1\over c_x^2}\cos^2\phi.
\end{equation}
Thus for a triaxial halo of mass $M_{vir}$ at a fixed redshift $z$, its lensing
signal depends on the line-of-sight direction. The 
statistical uncertainties of the axial ratios ($c_x$ and $c_y$) and of the 
concentration parameter ($c_e$) 
also contribute to the dispersion of lensing signals.

To demonstrate the dispersion caused by different viewing directions,
in Figure 3 we show the simulation result (solid line) of 
the distribution of the associated peak $\nu$ values of a halo of 
$M=3.9\times 10^{14}h^{-1}\hbox{M}_{\odot}$
at $z=0.48$ viewed along different line of sights. Also shown in the
plot (dashed line) is the smoothed central $\nu$ distribution 
expected from a triaxial halo of this mass and redshift with 
its axial ratios determined from the real mass distribution of the halo. 
The concentration parameter $c_e$ is taken to be the average value
(Jing \& Suto 2002). It is seen that the lensing signal has 
a large dispersion, and the triaxial model does explain a large 
portion of it at high end. 
%The good agreement between the solid and the dashed line indicates that
%for this particular case, the dispersion of the lensing signal
%can be fully explained by the triaxial mass distribution of the halo itself,
%and other effects do not play significant roles.

We now analyze the distribution of $\nu$ expected from the triaxial model for a given $\nu_{NFW}$. Because $\nu_{NFW}$ depends both on the mass
and on the redshift of a halo, halos of different masses at different redshifts
can have same $\nu_{NFW}$. Thus the conditional distribution is
%$f(\nu)=p(\nu |\nu_{NFW})$ is
\begin{equation}
f(\nu)=p(\nu|\nu_{NFW})=\frac{\int_0^1dz\frac{dV}{dzd\Omega}n[M_{\rm vir}(\nu_{NFW},z),z]p\,[\nu|\nu_{NFW};M_{\rm vir}(\nu_{NFW},z),z]dz}{\int_0^1dz\frac{dV}{dzd\Omega}n[M_{\rm vir}(\nu_{NFW},z),z]},
  \end{equation}
where $n(M_{\rm vir},z)$ is the mass function of dark matter halos. 
The probability function $\hbox {        }$          
$p[\nu|\nu_{NFW};M_{\rm vir}(\nu_{NFW},z),z]$ depends 
on the distributions of $\theta$, $\phi$, $c_x$, $c_y$ and $c_e$. Specifically,
\begin{eqnarray}
p[\nu|\nu_{NFW};M_{\rm vir}(\nu_{NFW},z),z]d\nu= && \bigg \{\int dc_x \int d(c_x/c_y)\int dc_e \int d\theta \bigg ({\partial \nu \over \partial \phi}\bigg )^{-1}_{\theta, c_x,c_x/c_y, c_e} \nonumber \\
&& p(c_x)p[(c_x/c_y)|c_x]p(c_e)p(\theta)p(\phi)\bigg \} d\nu
\end{eqnarray}
%\begin{equation}
%p(\nu|\nu_{NFW};M_{\rm vir},z)d\nu=\bigg \{\int dc_x \int d(c_x/c_y)\int dc_e
%\int d\theta \bigg ({\partial \nu \over \partial \phi}\bigg )^{-1}_{\theta, c_x,c_x/c_y, c_e}p(c_x)p[(c_x/c_y)|c_x]p(c_e)p(\theta)p(\phi)\bigg \} d\nu,
%\end{equation}
where the distributions on $c_x$, $c_x/c_y$, and $c_e$
are taken from Jing and Suto (2002) [see also Oguri et al. (2003; 2004)]. 
The mass function of Jenkins et al. (2001) is adopted.  

In Figure 4, we show the $\nu$ distribution for 
$\nu_{NFW}=4.5, 5$ and $6$, 
%$\nu_{NFW}=4.5\pm 0.05, 
%5\pm 0.05$ and $6\pm 0.05$, 
respectively. The solid lines are the results from 
$\kappa$ maps of three simulation runs. The dashed lines are the results
of equation (22). The dash-dotted lines are the results for spherical halos
taking into account the uncertainties of the concentration parameter.
It is clearly seen that the dispersion solely from the distribution of 
the concentration parameter (dash-dotted lines) is small in comparison 
with the large dispersion shown in simulation data (solid lines). 
The triaxiality contributes additional dispersions, giving rise to a
long tail at high $\nu$ side. 
%It is notably seen, however, that the 
%range of $\nu$ from simulations is broader than that 
%predicted by the triaxial model, especially at low end of $\nu$.
%Possible reasons for this are discussed in the following.
 
%Comparing the dashed and dash-dotted lines, we see that the triaxiality 
%of halos affects the lensing signals significantly, giving rise to a
%long tail at high $\kappa$ side. Confronted with simulation results,
%the triaxiality contributes a considerable 
%part of the high-end $\nu$ (or $\kappa$) dispersion, 
%especially for large $\nu_{NFW}$. The lower
%side extension, however, is notably wider than that from
%the triaxial model, which we will discuss further in this paper. 

The high-end dispersion due to the triaxiality of dark matter halos
has important effects on modeling the selection function for weak 
lensing cluster surveys. Since the lensing signal from a triaxial
cluster depends not only on its mass and redshift but also on the viewing
direction, the selection criteria corresponding to a 
detection limit $\kappa_{lim}$ is no longer simply a mass limit 
$M_{lim}(z)$ for each redshift $z$. Instead, for each $M_{vir}$, there is
a probability $p(M_{vir},z;\kappa_{lim})$ that its $\kappa$ can be higher
than $\kappa_{lim}$. Eq.(23) is the basic equation to obtain
$p(M_{vir},z;\kappa_{lim})$. One of our future studies
is to analyze $p(M_{vir},z;\kappa_{lim})$ and further the weak lensing
cluster abundances at different redshifts for different cosmologies. 

It is seen from Figure 4 that the triaxial model
explains the dispersion better for more massive clusters. 
It is notably seen, however, that the
range of $\nu$ from simulations is broader than that
predicted by the triaxial model, especially at the low end of $\nu$.
For relatively weak detections ($\nu_{NFW}=4.5$), the differences
between the results from the model predictions and the simulations 
show up at both low and high $\nu$ values.
Possible reasons for these are discussed in the following.

The differences between the simulation
results and the triaxial model predictions can be attributed
to two effects. One is the complex mass distribution of 
individual clusters that cannot be well approximated by the 
triaxial model. The other is the projection effect of uncorrelated 
structures along the line of sight.
To see the relative importance of the two, we generate lensing
surveys for matched halos only. Specifically, only particles 
that belong to the matched halos are kept and all the other particles
are removed from the simulation. In this way, we isolate the effect 
of the complexity of individual clusters. The lensing signal from
this artificial survey is referred to as $\kappa_{single}$ 
(or equivalently $\nu_{single}$).

Figure 5 shows the scatter plots of $\nu_{single}$ vs. $\nu_{peak}$ for 
$\nu_{NFW}=4.5, 5$ and $6$, respectively, where $\nu_{peak}$ is
the peak $\nu$ value of matched halos from full simulations. 
Good correlations between the two are clearly seen. 
This association indicates that for these peaks, their $\kappa$ values
are dominantly determined by the properties of their matched halos.
The plots also reveal that the dispersion toward high $\nu_{peak}$
is larger for smaller halos (lower $\nu_{NFW}$). This shows, as expected,
that the line-of-sight projection effects are relatively 
stronger for smaller halos. Quantitatively, we calculate the
quantity $<[(\nu_{peak}-\nu_{single})/\nu_{single}]^2>^{1/2}$ to
estimate the strength of the projection effect.
The values are $0.25$, $0.18$ and $0.15$ for $\nu_{NFW}=4.5, 5$ and 
$6$, respectively. In the plots, the triangles denote 
clusters with more than one $\nu$ peak associated with them.
These clusters must contain large substructures. 
Note that we identify the $\nu$ peak closest to the central 
position of a halo as its matched peak. For clusters with large substructures,
the closest $\nu$ peak is likely generated only by a substructure. 
Thus the value of matched $\nu_{peak}$ 
underestimates the lensing signal from the full cluster.
We indeed see that the triangles concentrate on the small $\nu_{peak}$ part. 
This explains the lower-end extension of the distribution
of $\nu_{peak}$ in Figure 4.

Now let us compare $\nu_{single}$ with $\nu_{tri}$ calculated from 
the triaxial model in which we determine the axial ratios and the 
orientation of a cluster from its real mass distribution. 
The results are shown in Figure 6. The symbols
are the same as in Figure 5. First we do see correlations 
between $\nu_{single}$ and $\nu_{tri}$, with the associations better
for single-peak clusters (crosses). It is noted, however,
that the triaxial model predicts 
narrower scatters than $\nu_{single}$ for relatively small clusters.
For large clusters with $\nu_{NFW}=6$, the spread of $\nu_{tri}$
is about the same as that of $\nu_{single}$.
Secondly, as expected, most of the triangles
are in the lower right part of the plot. But there are several
triangles scattered at the upper left part. We check individual cases,
and find that multiple substructures along the line of sight can produce
a higher lensing signal than that predicted by the smoothed triaxial 
mass distribution. 
%Large scatters of $\nu_{single}$ are dominantly caused by the complex
%substructures within clusters.

We caution that the simulations we use are relatively small-sized
($100h^{-1}\hbox{ Mpc}$), and the number of large clusters
is small. As we stack simulation boxes together to generate lensing maps,
the same cluster can be surveyed many times along different directions. 
For $\nu_{NFW}=4.5, 5$, and $6$, the number of matched peaks
from lensing maps is, respectively, about $400$, $300$, and $200$. However,
the corresponding total number of clusters involved 
in our analyses is only $26, 23$ and $14$. The number of
times that the most frequently surveyed clusters appear in our lensing 
maps can account for about $25\%$ of the total number of matched peaks. 
Thus our statistical results can be 
influenced significantly by the characteristics of only a few clusters.
Similar analyses on larger simulations are desired for more robust 
statistical results. 
%Of these, the respective number of clusters
%with apparent substructures shown in lensing maps is $11, 12$
%and $10$, with the corresponding fraction
%$42\%$, $52\%$ and $71\%$. 

We further do case-studies one by one. Of the 
respective $26, 23$ and $14$ clusters for $\nu_{NFW}=4.5, 5$, and $6$, 
the number of clusters with apparent substructures shown in 
lensing maps is $11, 12$ and $10$, with the corresponding fraction 
$42\%$, $52\%$ and $71\%$. 
For these clusters, their lensing signals
are mostly lower than those predicted by the triaxial model
as explained earlier. A specific example is plotted in Figure 7.
The $z$-projected mass distribution of this cluster is shown in Figure 8.
Large subhalos are clearly seen. In fact, in this case,
the cluster is in its early formation stage, and different substructures
are just starting to merge.
For those clusters
without apparent substructures, most of them can be 
well described by the triaxial model. However, multiple hidden substructures 
along the line of sight can generate lensing signals higher than
those from the smoothed triaxial model. 
An example is shown in Figure 9. For this cluster, the 
expected lensing signal from the spherical NFW mass distribution is
$\nu_{NFW}=6$. Taking into account the triaxiality of the 
mass distribution, the predicted lensing signals
largely agree with that calculated directly from the real mass
distribution of the cluster. For the extreme point where 
$\nu_{single}$ is much higher than $\nu_{tri}$, the mass distribution 
along this particular line of sight is shown in Figure 10. The upper panel
is the real distribution and the lower one is from the 
smoothed triaxial model. Multiple major peaks are clearly seen in 
the upper panel, which explains the deviation of $\nu_{single}$ 
from $\nu_{tri}$. Notice that when applying the triaxial model
to individual clusters, we adopt the mean concentration parameter 
from Jing and Suto (2002) for a given set of axial ratios measured 
from the mass distribution of a cluster. The real concentration parameter can
fluctuate around the mean. We do find that, in a few cases ($5$ or $6$ out of 
$23$ clusters with $\nu_{NFW}=5$), 
the real mass distribution is much flatter than that of the
triaxial model with the mean concentration parameter, which 
results higher $\nu_{single}$ than $\nu_{tri}$.

The above analyses were done for the smoothing radius 
$\theta_G=1\hbox{ arcmin}$. To see the effects of $\theta_G$, 
we also perform studies with $\theta_G=2\hbox{ arcmin}$. Figure 11
shows the dispersions for $\theta_G=2\hbox{ arcmin}$. Only the results
from simulations (solid lines) and from the triaxial model (dashed lines)
are shown. It is seen that 
the differences between the simulation results and the predictions
of the triaxial model are larger than those for $\theta_G=1\hbox{ arcmin}$.
Quantitatively, for $\theta_G=1\hbox{ arcmin}$ and $\nu_{NFW}=6$, 
the dispersions are $\sigma_{simu}\approx 1.5$ and $\sigma_{tri}\approx 0.9$, 
respectively. The corresponding numbers for $\theta_G=2\hbox{ arcmin}$ are
$\sigma_{simu}\approx 1.8$ and $\sigma_{tri}\approx 0.6$. In Figure 12, we show
the scatter plots of $\nu_{single}$ and $\nu_{peak}$. First, as expected,
the fraction of triangles is significantly less than that of 
$\theta_G=1\hbox{ arcmin}$. For $\theta_G=2\hbox{ arcmin}$, the fraction
is $9\%$, $19\%$ and $11\%$ for $\nu_{NFW}=4.5$, $5$, and $6$, 
respectively. For $\theta_G=1\hbox{ arcmin}$, the corresponding 
numbers are $23\%$, $32\%$, and $38\%$. The decrease of the effect
of substructures does not however reduce significantly 
the dispersion of $\nu_{peak}$
at low end for $\theta_G=2\hbox{ arcmin}$.   
Secondly, the correlations of the two 
can still be seen in Figure 12, but the tightness of the association
is worse than the case of $\theta_G=1\hbox{ arcmin}$.
For $\theta_G=2\hbox{ arcmin}$ and $\nu_{NFW}=6$, 
we have $\sigma[(\nu_{single}-\nu_{peak})/\nu_{single}]\approx 0.24$. 
For $\theta_G=1\hbox{ arcmin}$, 
$\sigma[(\nu_{single}-\nu_{peak})/\nu_{single}]\approx 0.15$.
The large dispersion reflects
the severe projection contamination for $\theta_G=2\hbox{ arcmin}$.
In Figure 13, we show the particle distribution at several lens planes
for a particular peak with $\nu_{peak}=4.5$. This peak appears in both cases of 
$\theta_G=1\hbox{ arcmin}$ and $\theta_G=2\hbox{ arcmin}$. The
corresponding primary cluster of the peak is located at plane $11$.
The three circles in each panel are for angular scales of 
$1\hbox{ arcmin}$, $2\hbox{ arcmin}$ and $3\hbox{ arcmin}$, respectively.
It is clearly seen that within the circle of $1\hbox{ arcmin}$, 
the cluster itself is the dominant contributor to the lensing effect,
and the projection effect from other planes is relatively weak. 
Within the circle of $2\hbox{ arcmin}$, however, besides the cluster,
the mass distribution on planes $7, 13, 31, 33$ and $39$ 
contributes significantly to the final lensing effect. Specifically,
$(\nu_{peak}-\nu_{single})/\nu_{single} \approx 0.37$ and 
$0.8 $ for $\theta_G=1\hbox{ arcmin}$
and $\theta_G=2\hbox{ arcmin}$, respectively.
We note that the projection effect causes large dispersions
at both high and low ends. The joint effects of the 
stronger projection contamination and the reduction of substructures
make the low-end dispersion not change greatly in comparison with that 
of $\theta_G=1\hbox{ arcmin}$. At the high end, however, 
the dispersion is significantly larger for $\theta_G=2\hbox{ arcmin}$
than for $\theta_G=1\hbox{ arcmin}$.
Thus our analysis indicates that it seems too large 
to use $\theta_G=2\hbox{ arcmin}$ in weak lensing
cluster surveys.

\section{Discussion}

Our study shows that for clusters of galaxies, their peak
lensing signals appear much more complex than those predicted 
by the spherical mass distribution in which the total mass and the redshift
of a cluster fully determine the strength of its lensing effects. 
Therefore the spherical model can introduce large errors 
to the prediction on the abundance of lensing-selected clusters.
Those errors in turn will significantly bias the 
determination of cosmological parameters from future weak lensing 
cluster surveys.

Hamana et al. (2004) attribute the complexity to the variation of 
the concentration parameter of the spherical NFW profile and the 
line-of-sight projection effect of uncorrelated structures.
However, our analyses with $\theta_G=1\hbox{ arcmin}$
reveal that for relatively large clusters 
relevant to weak lensing cluster surveys, their lensing effects
are mainly determined by the characteristics of individual clusters,
and the line-of-sight projection effects play a minor role. 
For individual clusters, the triaxiality of their mass distribution 
is important and has to be taken into account in theoretical 
analyses on their expected lensing effects. In terms of 
the dispersion shown in Figure 4, the triaxiality contributes
a considerable part of the high-end tail especially for massive clusters. The 
low-end extension is mainly attributed to the existence of large substructures
in clusters of galaxies. Multiple substructures along a line of sight
can also generate higher lensing signals than those predicted
by the smoothed mass distribution. Thus substructures also contribute,
to a certain level, to the high-end dispersion. 
With $\theta_G=2\hbox{ arcmin}$, however, the projection effect is significant. 
Thus in terms of cluster detections with weak lensing surveys,
the smoothing radius $\theta_G=1\hbox{ arcmin}$ is preferable
to $\theta_G=2\hbox{ arcmin}$. 

Note that our analyses are based on the $\kappa$ maps generated under
the Born approximation. Therefore we may underestimate the projection effect
because we integrate the weighted density fluctuations
along the unperturbed light path. 
%Although the Born approximation
%is expected to be well valid in the weak lensing regime we are 
%interested in, full ray tracing should be done to measure the 
%projection effect more precisely. 
In our analysis with $\theta_G=1\hbox{ arcmin}$, we find that the projection effect
accounts for about $0.16$ in the total dispersion of $\sim 1.39$ of
$\nu_{peak}-\nu_{NFW}$ for $\nu_{NFW}=5$. 
Hamana et al. (2004) quoted a value of $\sim 0.5$ from the projection effect
estimated from the contribution of large-scale structures
uncorrelated with halos (Hoekstra 2001). While the number
$0.5$ is not certain and the underestimate by the Born approximation
in the weak lensing limit may not be as large as the numbers ($0.16$ vs. $0.5$) suggest, 
it is indeed desirable to quantitatively analyze the
projection effects by full ray tracing simulations.

In our analysis, we assume a fixed source redshift. For a real lensing survey, 
source galaxies generally have a redshift distribution, which should be taken into account
in modeling the lensing effects of clusters expected from the survey. 

With the mass distribution and the statistics 
for the triaxial dark matter halos from Jing and Suto (2002), 
the effect of the triaxiality on lensing signals 
can be theoretically studied. Rather than simply mass selected, 
a much more complicated selection function for $\kappa$-limited lensing 
cluster surveys can be obtained.
On the other hand, it is less straightforward to include
substructures in theoretical analyses without numerical simulations.
However, recent high resolution simulations have been able to present
some statistics on substructures, e.g., their mass function 
and the spatial distribution (e.g. Gao et al. 2004; 
Natarajan \& Springel 2004). These, in principle, 
allow us to analyze the influence of substructures on 
lensing effects without time-intensive simulations. 

Weak lensing effects are powerful and clean probes to the 
distribution of dark matter. Cosmological studies on lensing-selected 
clusters avoid physical processes related to the intracluster gas 
that complicate the cosmological applications of X-ray and SZ-selected 
clusters considerably. Our investigation reveals the complexities of 
lensing effects related to the mass distribution of clusters, which however,
can be handled without fundamental difficulties. In a foreseeable future, 
weak lensing cluster surveys as well as cosmic shear observations
will contribute greatly to cosmological studies.

\acknowledgments 
We sincerely thank the referee for the constructive and detailed comments and 
suggestions. We gratefully thank Y. P. Jing for 
kindly providing us simulation data and 
for useful discussions. We also thank G. T. Zhu for his $\kappa$-map
algorithm.  
This research was supported in part by the National Science Foundation of China
under grants 10243006 and 10373001, by the Ministry of Science 
and Technology of China under grant TG1999075401, and by the Key Grant
Project of Chinese Ministry of Education (No. 305001).
%%%%%%%%%%%%%%%%%%%%%%%%%%%%%%%%%%%%%%%%%%%%%%%%%%%%
%%%%%%%%%%%%%%%%%%%%%%%%%%%%%%%%%%%%%%%%%%

%\end{document}

\begin{figure}
\plotone{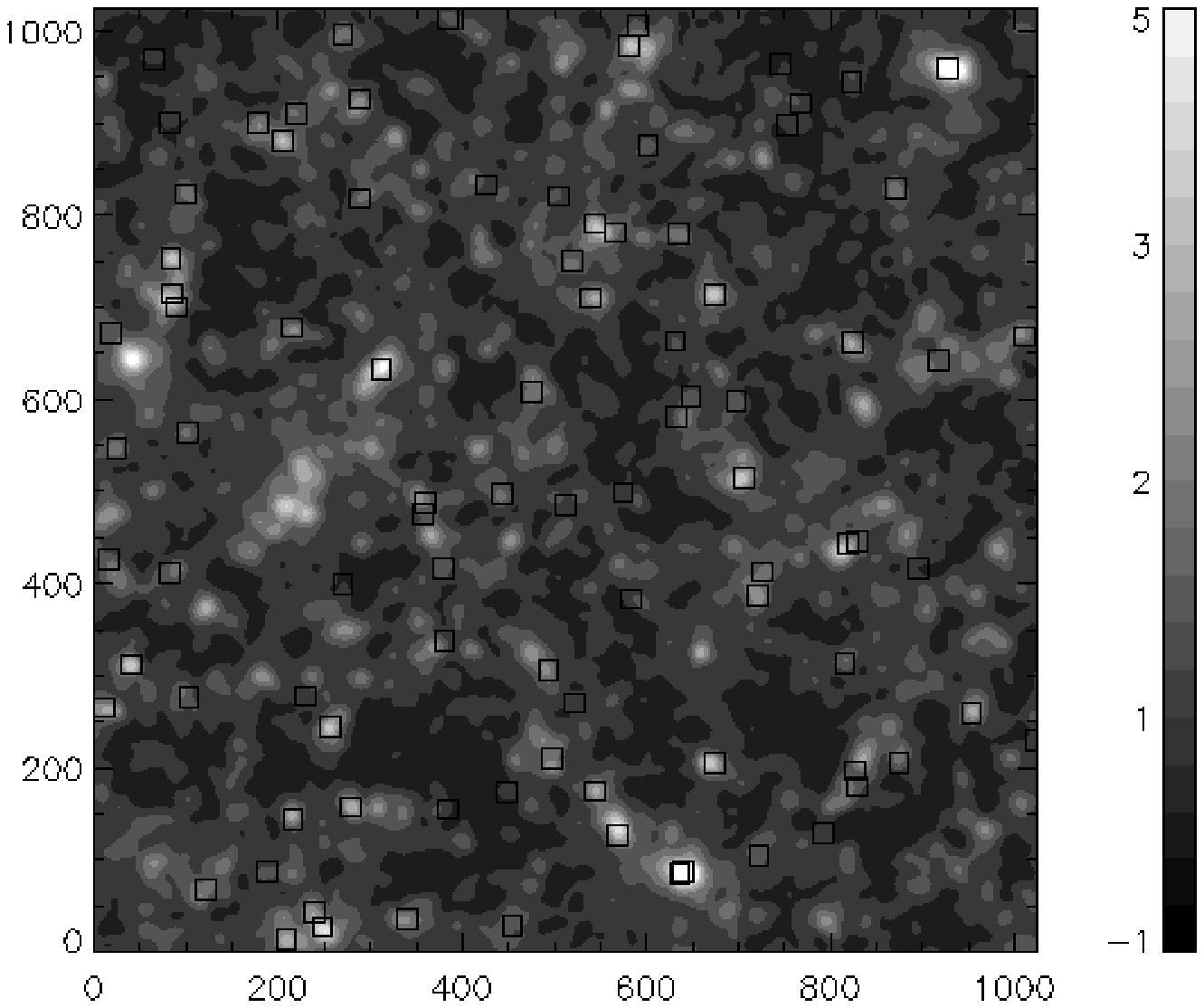} \caption{A $2^{o}\times 2^{o}$ $\kappa$-map
from a simulation. \label{yg1}}
\end{figure}

\begin{figure}
\plotone{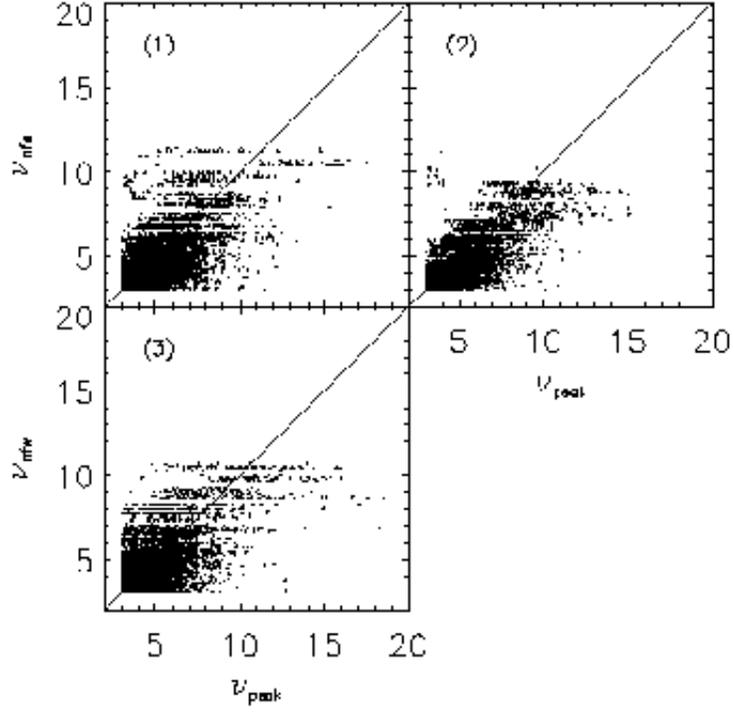} \caption{Scatter plots of $\nu_{peak}$ vs. 
$\nu_{NFW}$ for the three simulation runs, respectively. \label{yg2}}
\end{figure}

\begin{figure}
\plotone{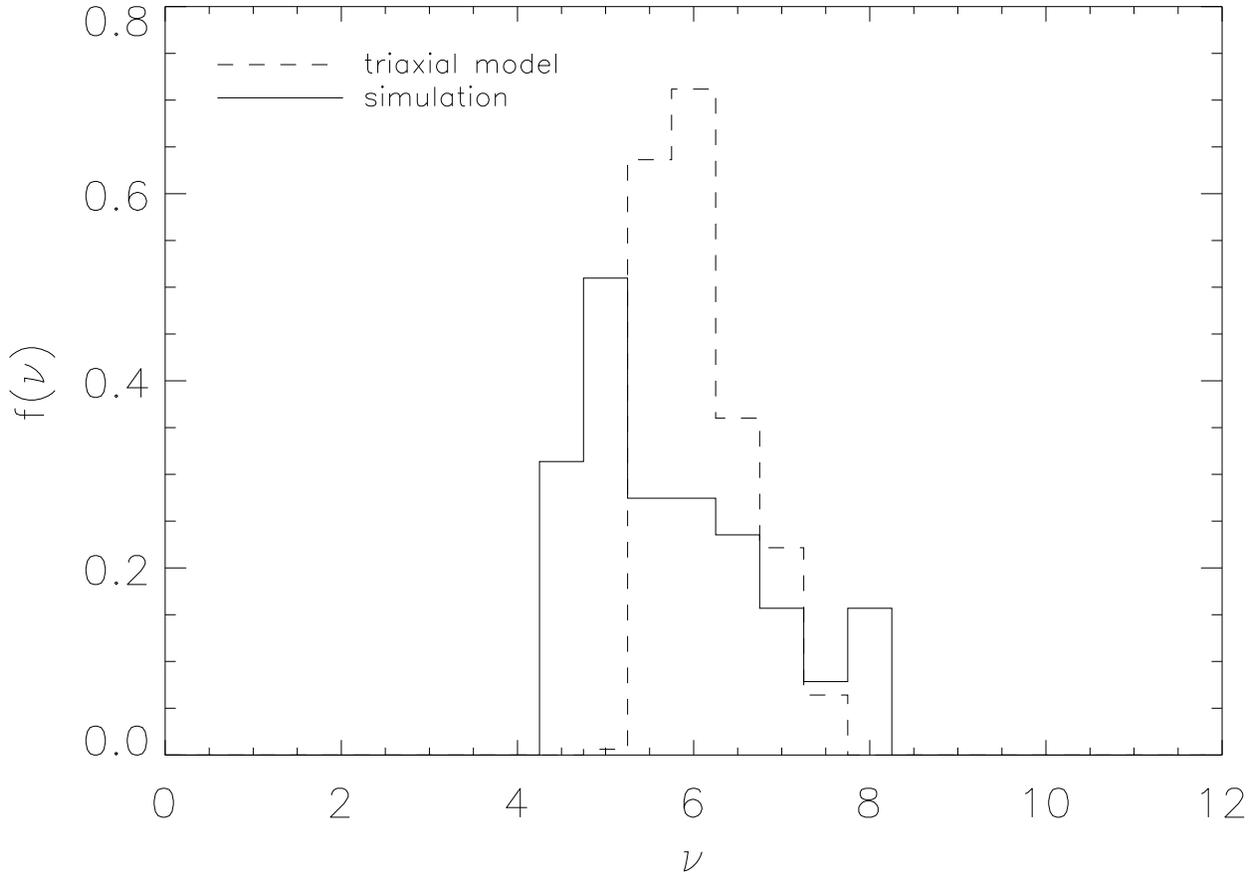} \caption{The distribution of $\nu$ for a particular
cluster with $M\approx 3.9\times 10^{14}h^{-1}\hbox{M}_{\odot}$
at redshift $z=0.48$ ($\nu_{NFW}=6$) viewed from different directions. 
The dashed line is the Monte-Carlo result of the triaxial model
for all the possible line-of-sight directions. 
The axial ratios are calculated from the mass distribution of the cluster.
The solid line is the result from simulations. \label{yg3}}
\end{figure}

\begin{figure}
\plotone{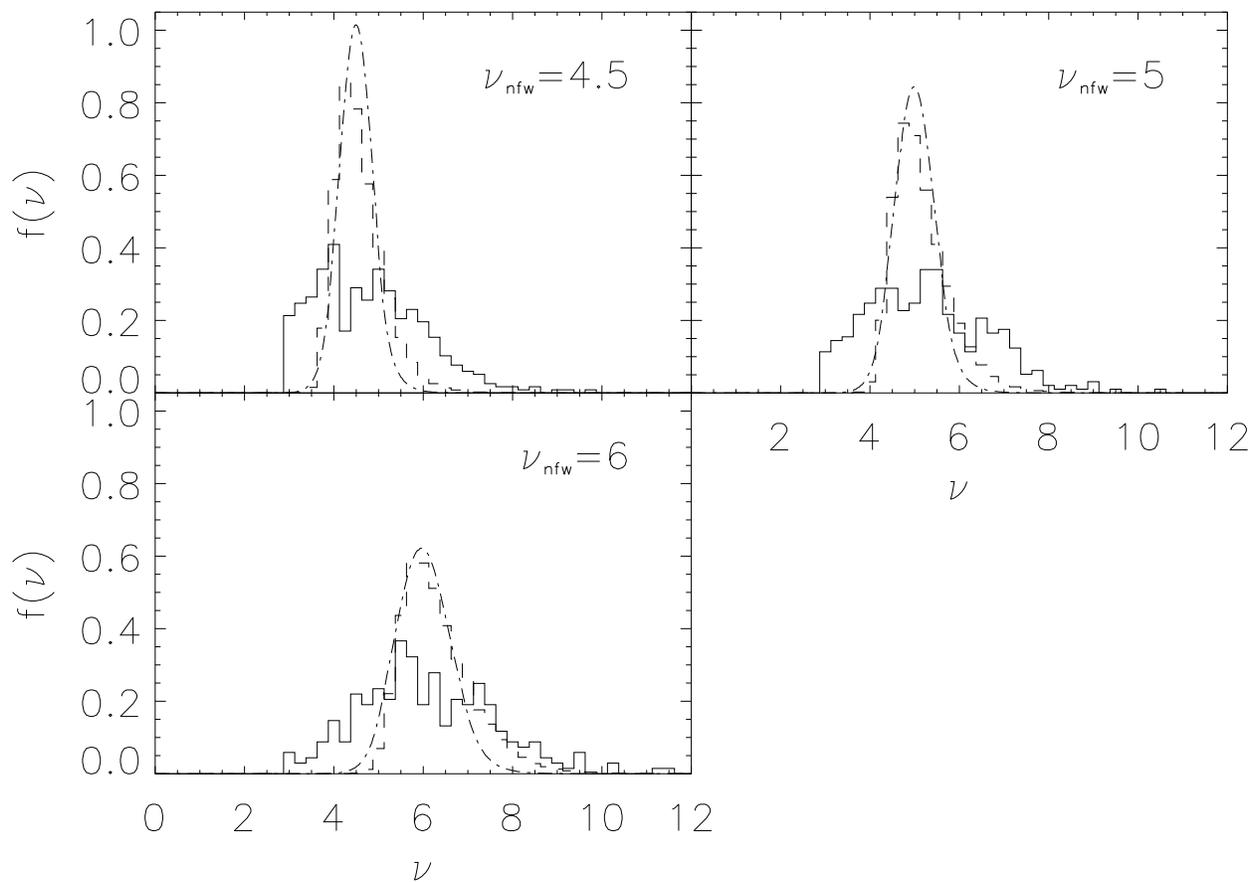}
 \caption{The distribution of $\nu$ for all the clusters with
 $\nu_{NFW}=4.5, 5$ and $6$, respectively. The solid line is the result 
 from $\kappa$ maps, the dashed line is the statistical result from 
 the triaxial model, and the dash-dotted line is the result of shperical
 NFW model taking into account the statistical variation of the concentration
 parameter.  \label{yg4}}
 \end{figure}

 \begin{figure}
 \plotone{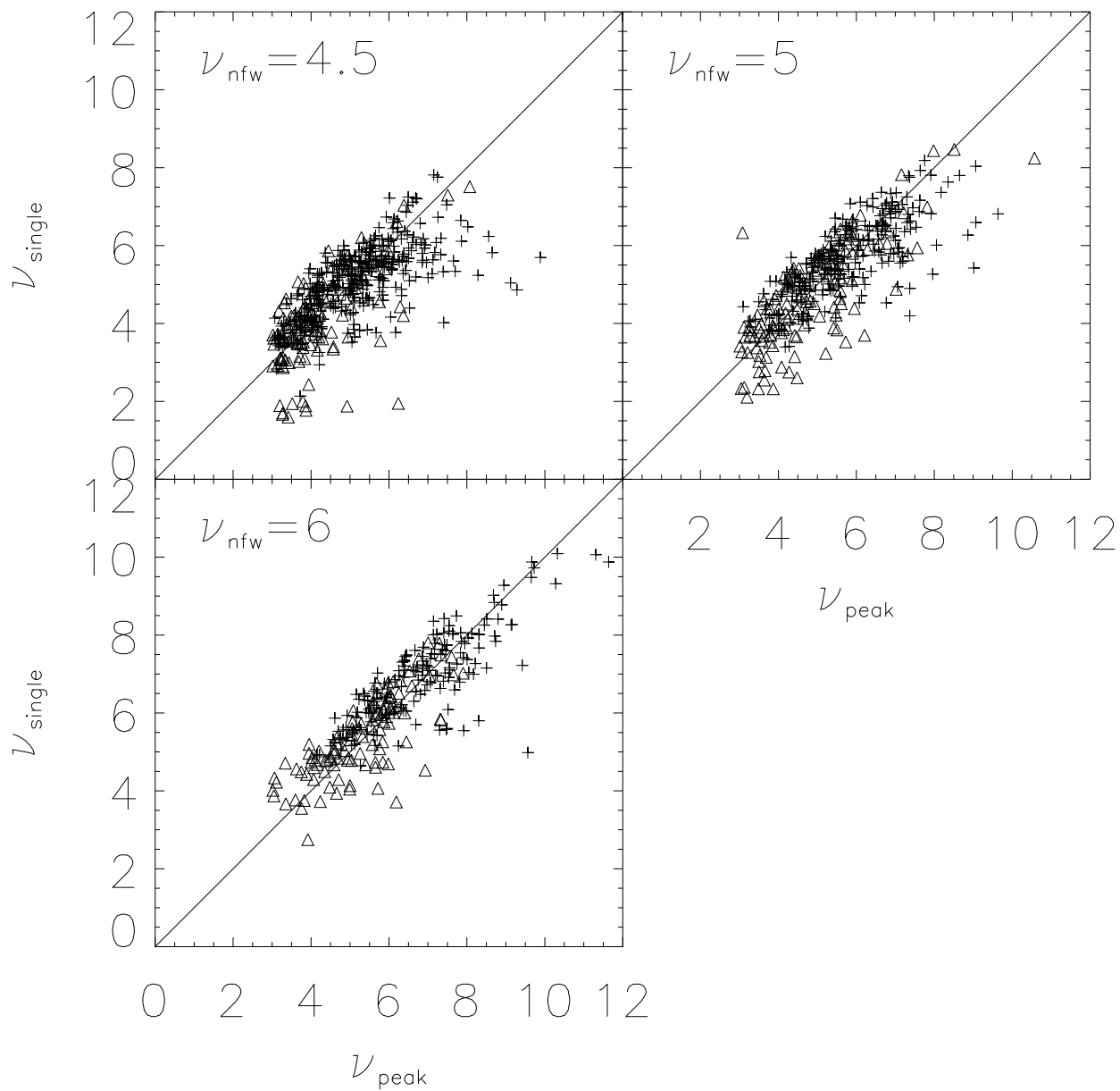}
  \caption{Scatter plots of $\nu_{peak}$ vs. $\nu_{single}$. The
  three plots are for $\nu_{NFW}=4.5, 5$ and $6$, respectively. 
  The triangles denote cases that multiple $\kappa_{single}$ peaks 
  appear within $2.88\hbox{ arcmin}$ around the center of the clusters.
  The crosses are for those that only one $\kappa_{single}$ peak
  exists within $2.88\hbox{ arcmin}$ around the center of the clusters.
  \label{yg5}}
  \end{figure}

\begin{figure}
\plotone{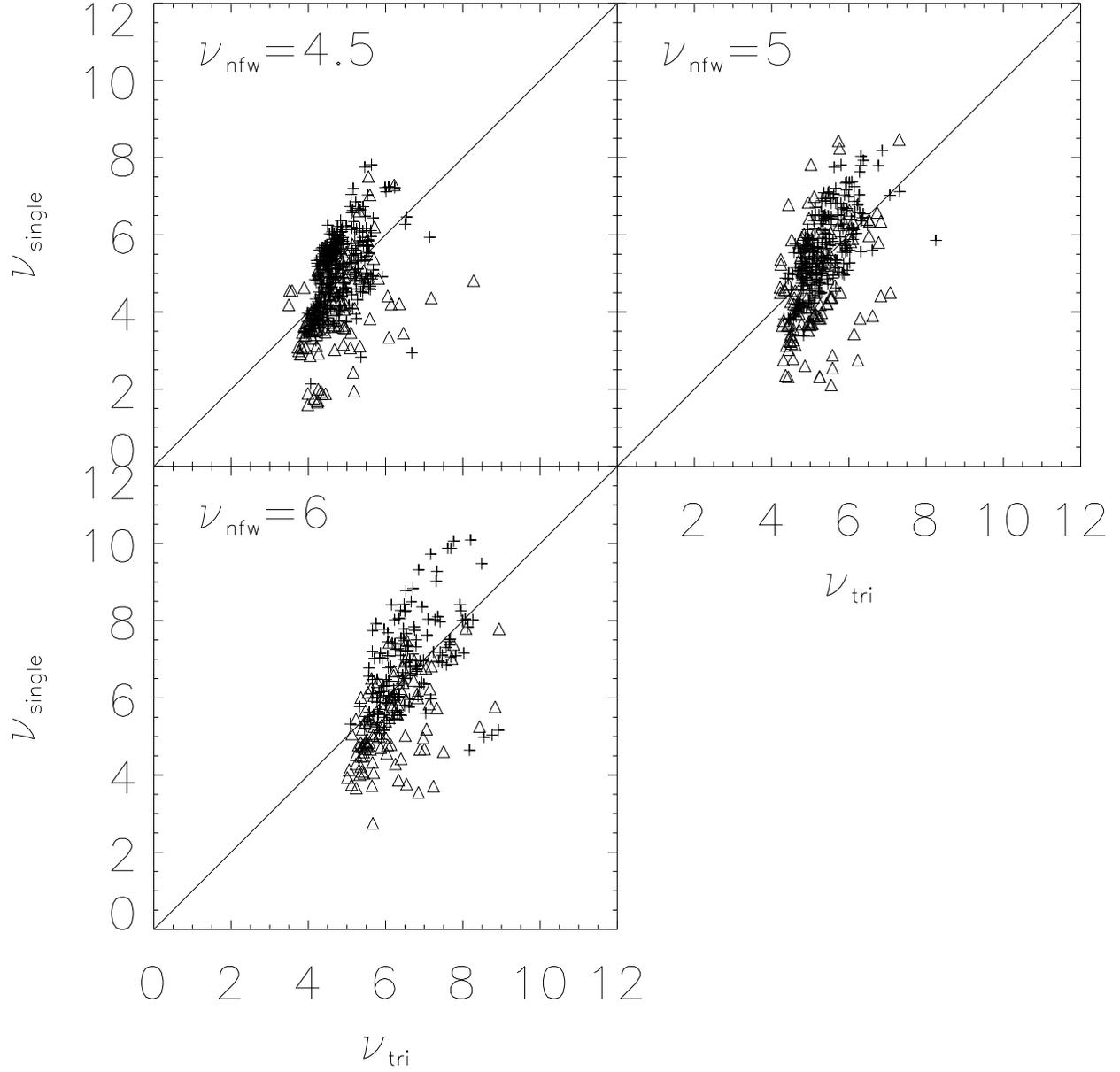}
\caption{Scatter plots of $\nu_{tri}$ vs. $\nu_{single}$. The symbols
are the same as in Fig5.
\label{yg6}}
\end{figure}

\begin{figure}
\plotone{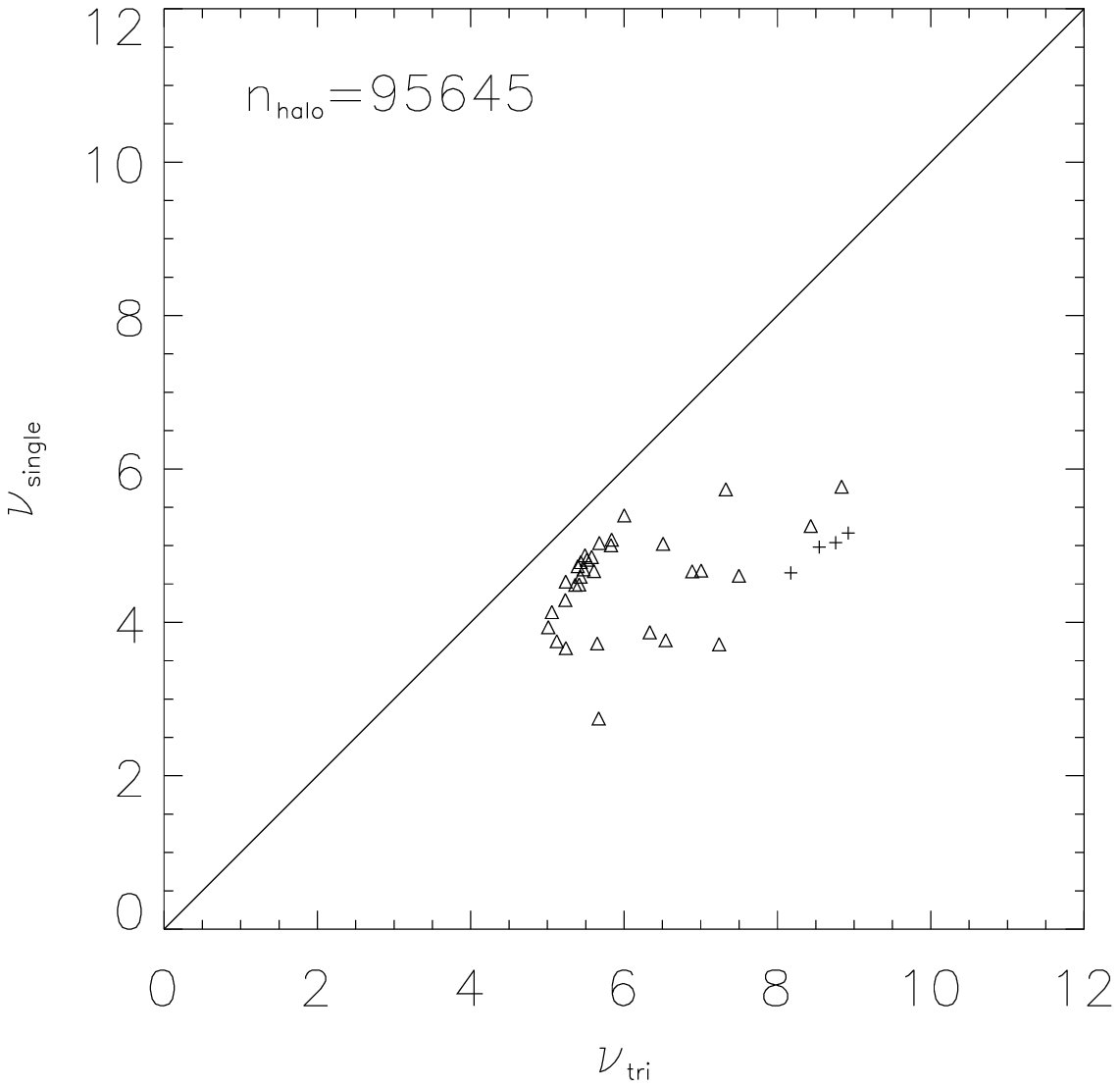}
\caption{The scatter plot of $\nu_{tri}$ vs. $\nu_{single}$ for
 a particular cluster '95645' with $\nu_{NFW}=6$.
 \label{yg10}}
 \end{figure}

 \begin{figure}
 \plotone{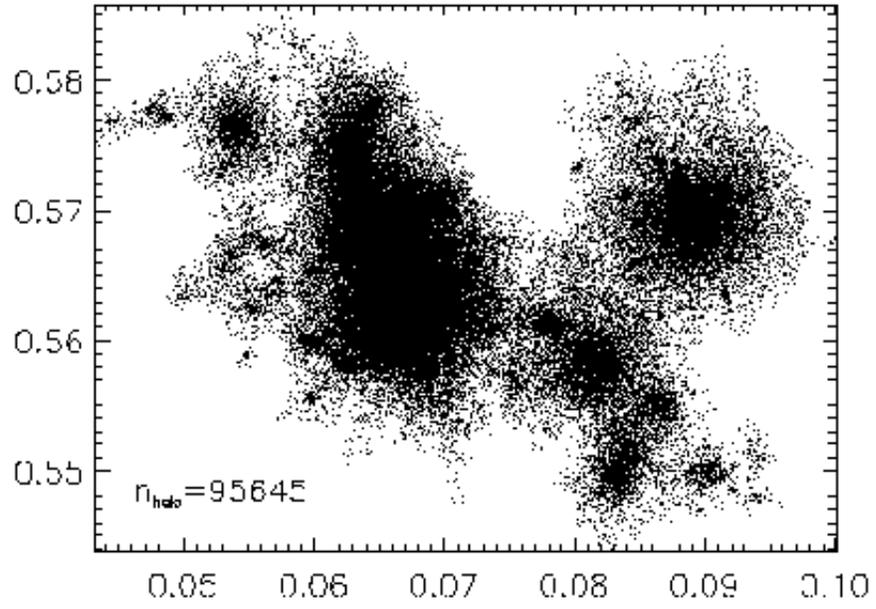}
 \caption{The projected particle distribution along $z$-direction
 for cluster '95645'.
 \label{yg9} }
 \end{figure}

\begin{figure}
\plotone{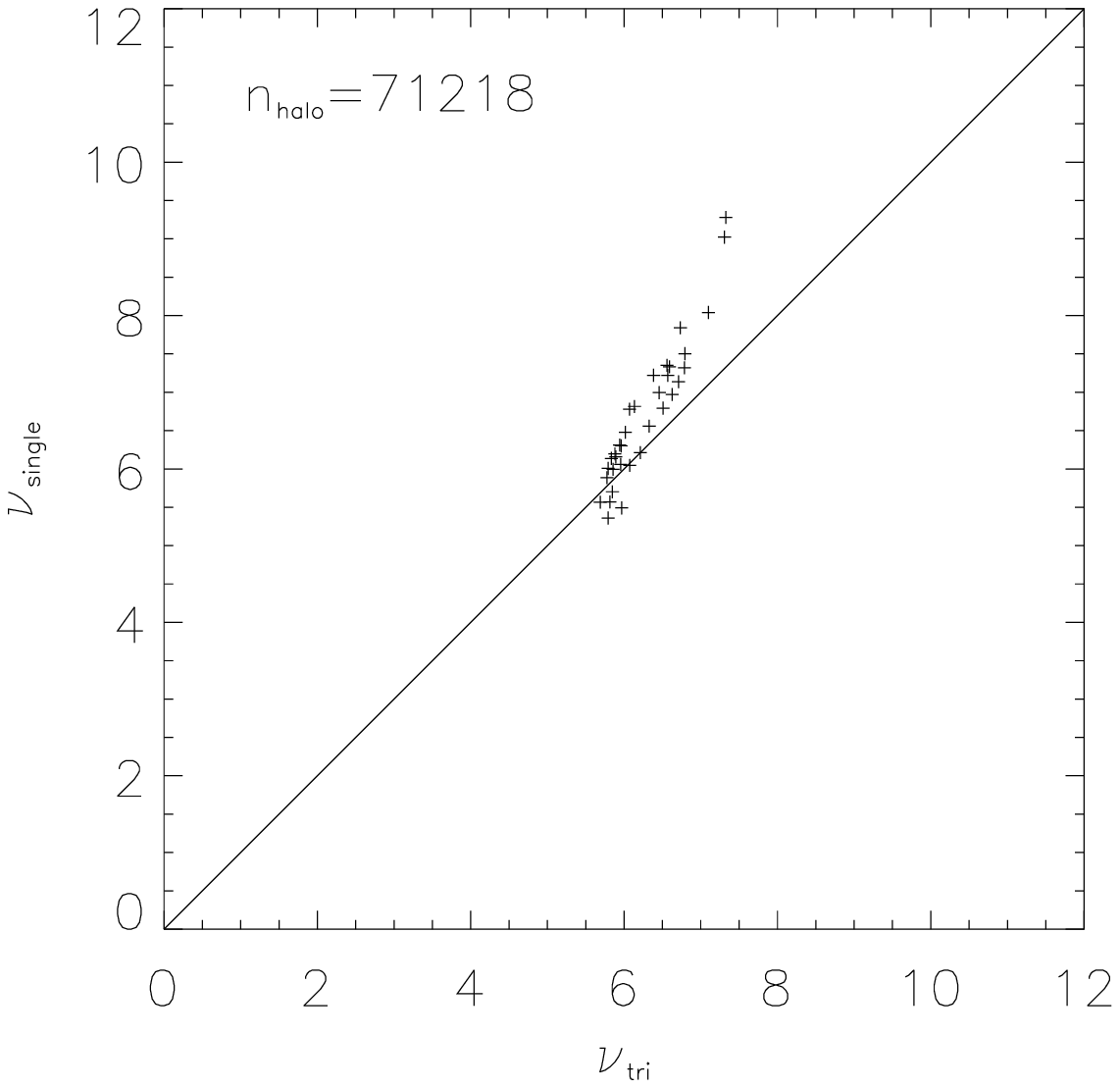}
 \caption{The scatter plot of $\nu_{tri}$ vs. $\nu_{single}$ for 
 a particular cluster '71218' with $\nu_{NFW}=6$. 
 \label{yg7}}
 \end{figure}

 \begin{figure}
 \plotone{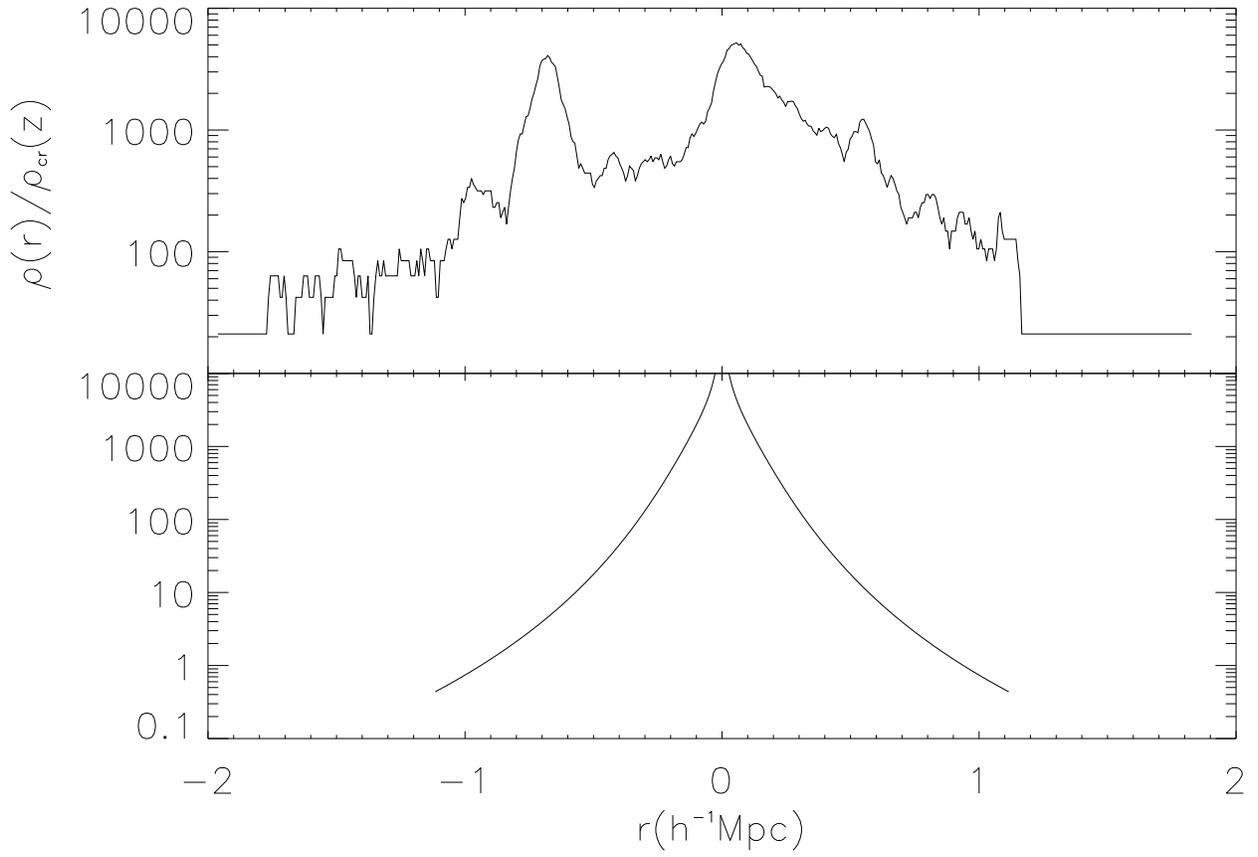}
  \caption{The mass density along the line-of-sight direction corresponding 
  to the highest point in Fig.9. The upper panel is the
  real mass distribution of the cluster, and the lower panel
  is the one from the smoothed triaxial mass distribution.
  \label{yg8} }
  \end{figure}

 \begin{figure}
  \plotone{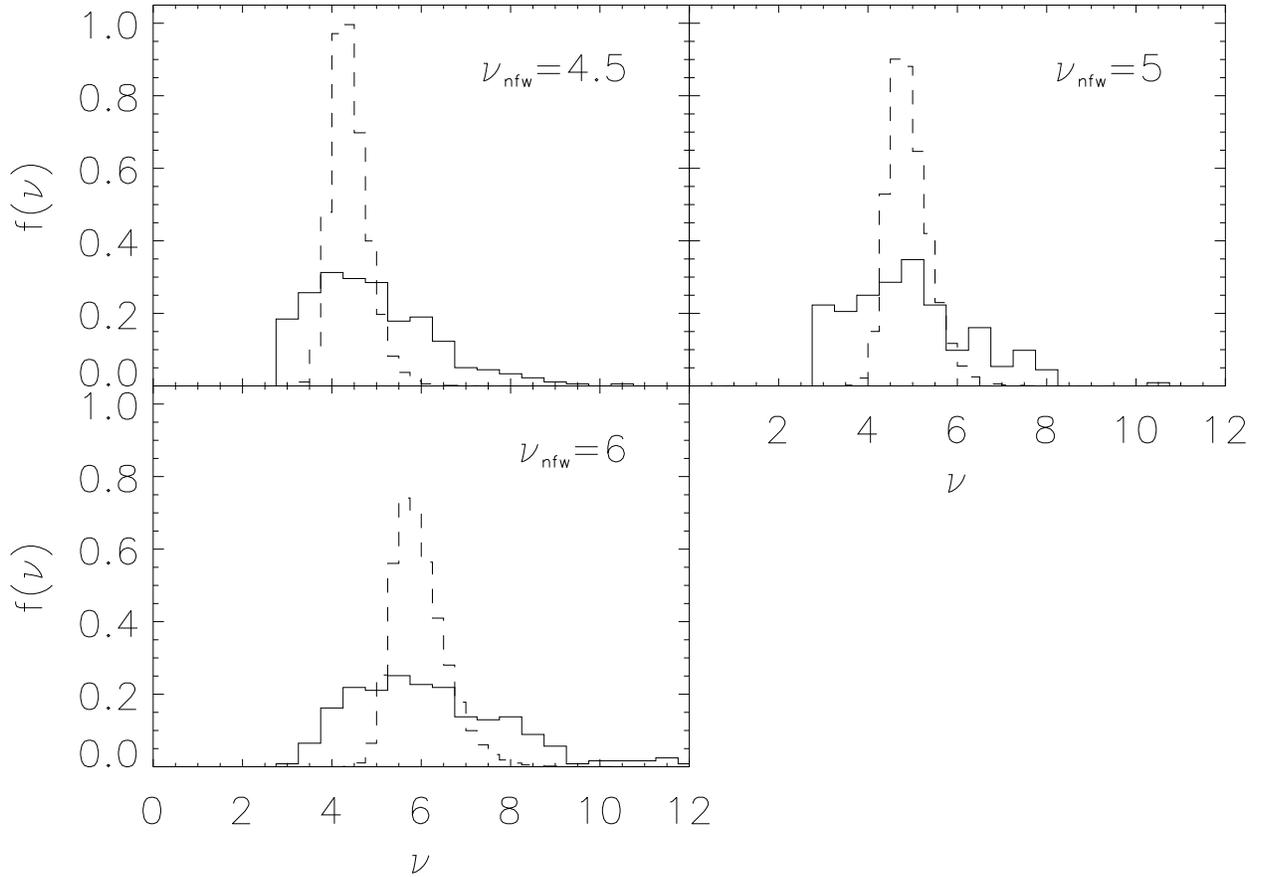}
    \caption{Same as Figure 4 but with $\theta_G=2\hbox{ arcmin}$. Solid lines
    are results from simulations and dashed lines are the predictions
    from the triaxial model.
  \label{yg8} }
  \end{figure}

 \begin{figure}
  \plotone{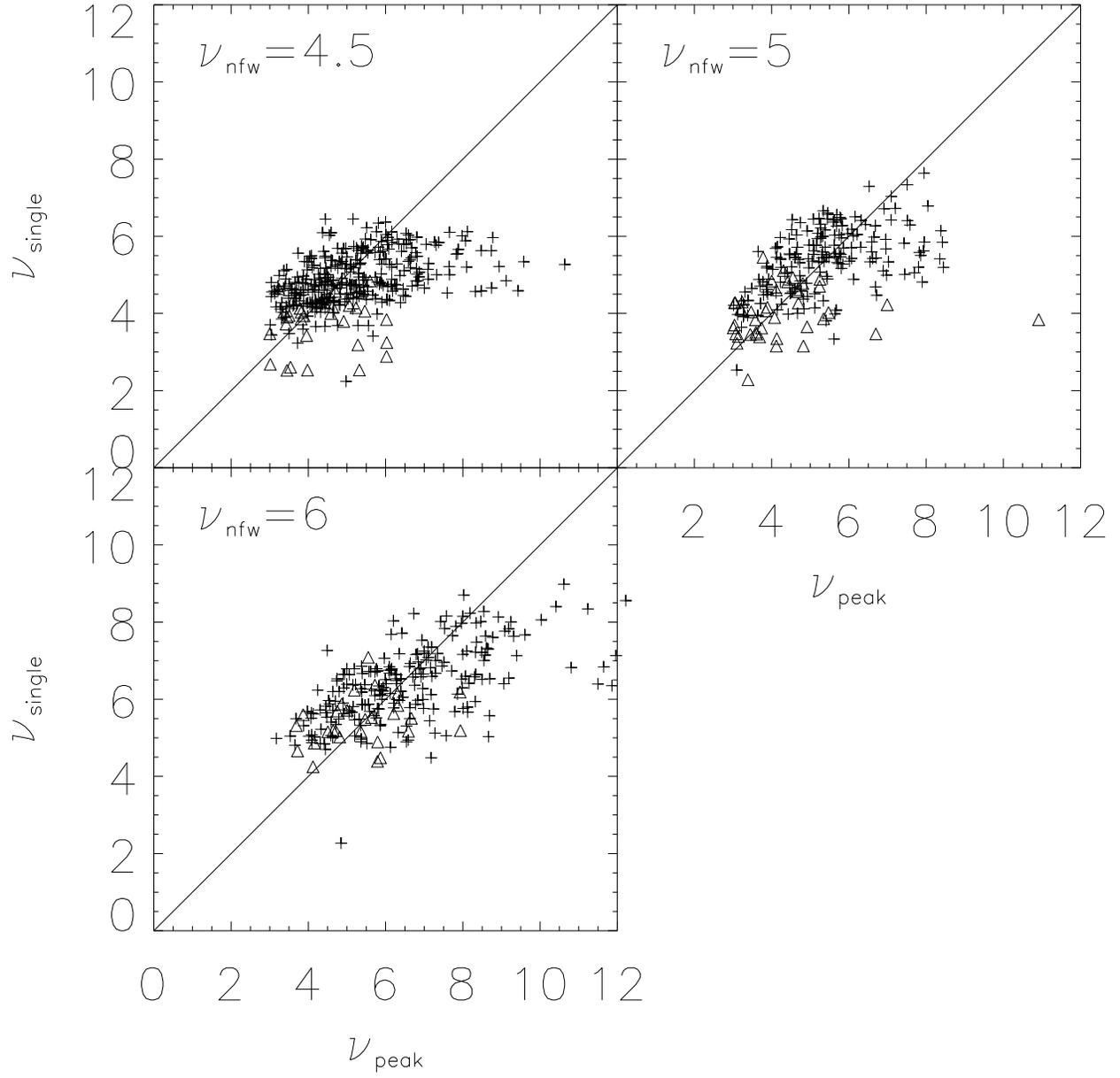}
    \caption{Same as Figure 5 but with $\theta_G=2\hbox{ arcmin}$.
  \label{yg8} }
  \end{figure}

 \begin{figure}
   \epsscale{0.8}
   \plotone{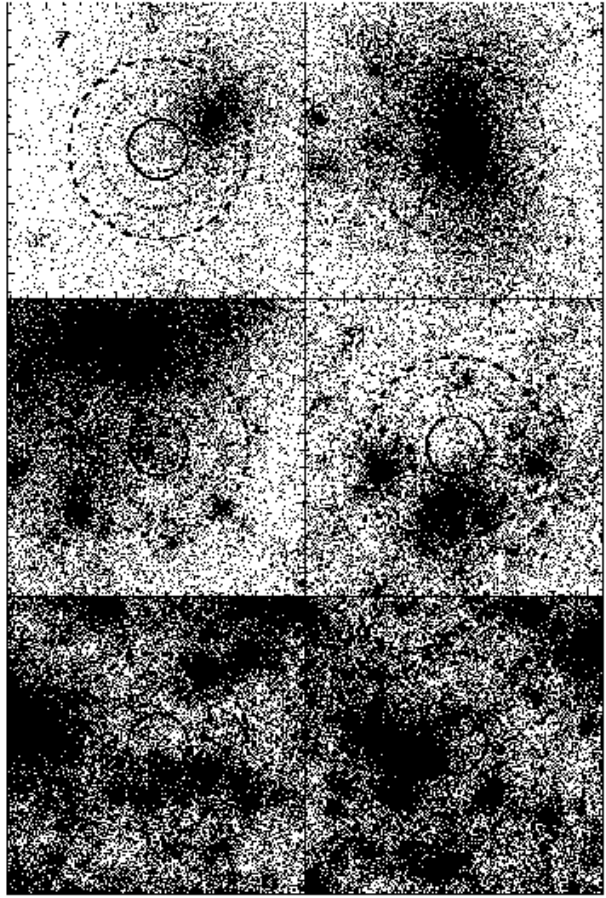}
   \caption{The particle distribution in lens planes $7, 11, 13, 31, 33$
   and $39$ for a particular peak with $\nu_{peak}=4.5$. The three
   circles from inner most to the outer most correspond to angular
   scales of $1\hbox{ arcmin}$, $2\hbox{ arcmin}$ and $3\hbox{ arcmin}$, 
   respectively.
   \label{yg8} }
   \end{figure}


\begin{thebibliography}{}
   \bibitem[Bahcall \& Bode (2003)]{}Bahcall, N., \& Bode, P. 2003, 
   \apj, 588, L1
   \bibitem[Bartelmann \& Meneghetti (2004)]{}Bartelmann, M., 
   \& Meneghetti, M. 2004, A\&A, 418, 413
   \bibitem[Bartelmann \& Schneider (2001)]{}Bartelmann, M., 
   \& Schneider, P. 2001, Phys. Rept. 340, 291
    \bibitem[]{} Carlstrom, J. E., Holder, G. P., \& Reese, E. D., 2002, \araa, 40, 643
   \bibitem[Fan \& Chiueh (2001)]{}Fan, Z. H., \& Chiueh, T. H. 2001, 
   \apj, 550, 547
   \bibitem[Fan \& Wu (2003)]{}Fan, Z. H., \& Wu, Y. L. 2003, 
   \apj, 598, 713
   \bibitem[Gao et al. (2004)]{}Gao, L., White, S. D. M., Jenkins, A., 
   Stoehr, F., \& Springel, V. 2004, \mnras, 355, 819
   \bibitem[Gavazzi et al. (2003)]{}Gavazzi, R., Fort, B., Mellier, Y., 
   Pello, R., \& Dantel-Fort, M. 2003, A\&A, 403, 11
   \bibitem[Haiman, Mohr \& Holder (2001)]{}Haiman, Z., Mohr, J., 
   \& Holder, G. 2001, \apj, 553, 545
   \bibitem[Haiman et al. (2004)]{}Haiman, Z., Wang, S., Hennawi, J. F., 
   May, M., Spergel, D.N., \& Tyson, J. A. 2004, American Astronomical
   Society Meeting 205, \#108.15
   \bibitem[Hamana, Colombi \& Mellier (2001)]{hcm01}Hamana, T., Colombi, S., 
   \& Mellier, Y. 2001, in XXXVth Moriond Astrophysics Meeting: Cosmological
   Physics with Gravitational Lensing. ed. J.-P. Kneib, Y. Mellier, M. Mon,
   \& T. Tran Thanh Van(France: EDP)
   \bibitem[Hamana, Takada \& Yoshida (2004)]{hty04}Hamana, T., Takada, M., 
   \& Yoshida, N. 2004, \mnras, 350, 893
   \bibitem[Hoekstra (2001)]{hoe01}Hoekstra, H. 2001, A\&A, 370, 743
   \bibitem[]{} Jain, B., Seljak, U., \& White, S. 2000, \apj, 530, 547
   \bibitem[]{} Jenkins, A. et al. 2001, \mnras, 321, 372
   \bibitem[Jing \& Suto (1998)]{}Jing, Y. P., \& Suto, Y. 1998, 
   \apj, 494, L5
   \bibitem[Jing, \& Suto (2002)]{J}Jing, Y. P., \& Suto, Y. 2002, \apj, 574, 538 
   \bibitem[Jing (2004)]{}Jing, Y. P. 2004, private communication
   \bibitem[Kaiser (1998)]{}Kaiser, N. 1998, \apj, 498, 26
      \bibitem[Lee \& Suto (2003)]{wwff1}Lee, J., \& Suto, Y. 2003,
\apj, 585, 151
       \bibitem[Lee \& Suto (2004)]{wwff36}Lee, J., \& Suto, Y. 2004,
\apj, 601, 599
    \bibitem[Ma (2003)]{}Ma, C. P. 2003, \apj, 584, L1
     \bibitem[Natarajan \& Springel (2004)]{}Natarajan, P., \& 
     Springel, V. 2004, \apj, 617, 13L
    \bibitem[Oguri, Lee \& Suto (2003)]{wwff33}Oguir, M., Lee, J., \& Suto, Y. 2003, \apj, 599, 70
    \bibitem[Oguri, Lee \& Suto (2004)]{}Oguir, M., Lee, J., \& Suto, Y. 2004, 
    \apj, 608, 1175
     \bibitem[Oguri \& Keeton (2004)]{ok}Oguri, M., \& Keeton, C.  2004,
\apj, 610, 6630
   \bibitem[Padmanabhan, Seljak \& Pen (2003)]{}Padmanabhan, N., Seljak, U.,
   \& Pen, U. L. 2003, New Astronomy, 8, 581
   \bibitem[Press \& Schechter (1974)]{}Press, W. H., \& Schechter, P. 1974, 
   \apj, 187, 425
   \bibitem[Reblinsky \& Bartelmann (1999)]{rb99} Reblinsky, K., \& 
   Bartelmann, M. 1999, A\&A, 345, 1
   \bibitem[]{}Rosati, P., Borgani, S., \& Norman, C. 2002, \araa, 40, 539
   \bibitem[]{}Sheth, R., \& Tormen, G. 1999, \mnras, 308, 119 
   \bibitem[Tereno, Dore, vanWaerbeke \& Mellier (2005)]{}Tereno, I., 
   Dore, O., van Waerbeke, L., \& Mellier, Y. 2005, A\&A, 429, 383
   \bibitem[vanWaerbeke (2000)]{vw00} van Waerbeke, L. 2000, \mnras, 313, 524
   \bibitem[vanWaerbeke, Mellier \& Hoekstra (2005)]{} van Waerbeke, L., 
   Mellier, Y., \& Hoekstra, H. 2005, A\&A, 429, 75
   \bibitem[Wang \& Fan (2004)]{wf04} Wang, Y.-G., \& Fan, Z.H. 2004, \apj, 
   617, 847
   \bibitem[White, vanWaerbeke \& Mackey (2002)]{wwm02}White, M., 
   van Waerbeke, L., \& Mackey, J. 2002, \apj, 575, 640
   \bibitem[Zhang (2004)]{}Zhang, T. J. 2004, \apj, 602, L5
\end{thebibliography}
\end{document}